\documentclass[12pt]{article}
\usepackage[pctex32]{graphicx}
\linethickness{0.4pt}

\begin{document}
\vskip 2.0 true cm	
\begin{center}
{\LARGE \bf Long-range Effects on the Pyroelectric Coefficient of Ferroelectric Superlattice}
\vskip 0.5 true cm
{\bf {\large Wen Dong $^{b},$ Dong-Lai Yao $^{b},$ Yin-Zhong Wu $^{b, c},$ Zhen-Ya Li $^{a, b}$}}
\vskip 0.2 true cm
{\it $^{a}$ CCAST World Laboratory, P.O.Box 8730,Beijing 100080, People's
Republic of China}\\
{\it $^{b}$ Department of Physics, Suzhou University, Suzhou 215006, China$^*$}\\
{\it $^{c}$ Department of Physics, Changshu College, Changshu 215500, China}\\
\end{center}

\begin{abstract}
Long-range effects on the pyroelectric coefficient of a ferroelectric superlattice consisting of two different ferroelectric materials are investigated based on the Transverse Ising Model. The effects of the interfacial coupling and the thickness of one period on the pyroelectric coefficient of the ferroelectric superlattice are studied by taking into account the long-range interaction. It is found that with the increase of the strength of the long-range interaction, the pyroelectric coefficient decreases when the temperature is lower than the phase transition temperature; the number of the pyroelectric peaks decreases gradually and the phase transition temperature increases. It is also found that with the decrease of the interfacial coupling and the thickness of one period, the phase transition temperature and the number of the pyroelectric peaks decrease.
\end{abstract}

\vspace{1.0cm}



$*$ mailing address in China\\

Electronic mail: zyli$@$suda.edu.cn\\

\newpage
{\bf{I. Introduction}}\\

Recently, more attentions were focused on the artificial ferroelectric superlattice. With the advancement of experimental techniques, the growth of $BaTiO_3 / SrTiO_3$ superlattice was achieved by molecular beam epitaxy [1,2], the $PbTiO_3 / PbZrO_3$ superlattice was fabricated by a multi-ion-beam sputtering technique [3], and the $(Sr, Ca) TiO_3 / (Ba, Sr) TiO_3$ superlattice was prepared by a pulsed laser ablation technique [4]. These superlattices all show some unique properties, which are different from those of the simple solids, and they may provide some new opportunities for ferroelectric material applications. Theoretically, many researches have been presented too. The Ginzberg-Landau phenomenological theory and the Transverse Ising Model (TIM) are used to study the spontaneous polarization, phase transition, dielectric and pyroelectric properties of the ferroelectric superlattice [5-10]. In experiment, scientists have observed that the long-range interaction have an important role on the ferroelectric materials [11,12]. In theory, Ma et al. [13] investigated the long-range interaction of a ferroelectric superlattice consisting of two kinds of ferroelectric materials by use of the Ginzberg-Landau phenomenological theory. They considered the effects of the long-range interaction on the polarization, the Curie temperature and the dielectric susceptibility of a ferroelectric superlattice. \\

As is well known, the most efficient pyroelectric detectors are made from ferroelectric materials, and the study of the pyroelectric effects of ferroelectric materials was active in experiment and theory [14-16]. Theoretically, Glinchuk et al. [17] investigated the size effects on pyroelectric coefficient in ferroelectric films. Pintilie et al. [18] obtained the equivalent pyroelectric coefficient of a pyroelectric bimorph structure. Xin et al. studied the pyroelectric properties of ferroelectric superlattice by use of TIM. But the effects of the long-range interaction on pyroelectric properties are not considered. In this paper, we investigate the effects of the long-range interaction on the pyroelectric properties in detail. We also study the effects of the interfacial coupling and the thickness of one period on the pyroelectric properties within the framework of the long-range interaction. It is found that the pyroelectric properties of the ferroelectric superlattice will be strongly influenced by the long-range interaction. Moreover, the phase transition temperature will increase with the increase of the strength of the long-range interaction.\\  

{\bf{II. Model AND Formulation}}\\

We consider a ferroelectric superlattice composed of two different ferroelectric materials, A and B, stacked alternately. In one period of the superlattice, we suppose the layer-number of each material is $N_\alpha  $ ($\alpha  = a, b$). The periodic boundary condition suggests that we only need to consider one period ($N = N_a  + N_b $). Each layer is defined on the x-y plane and pseudo spins site on these layers (see Fig. 1). The system can be described by the Ising model in the presence of a transverse field

\begin{equation}
H =  - \sum\limits_{i,j} {J_{ij} S_i^z S_j^z  - \sum\limits_i {\Omega _i S_i^x  - 2\mu E\sum\limits_i {S_i^z } } } ,
\end{equation} 
where  $\Omega _i$ is the transverse field at site $i$. $S_i^x$ and $S_i^z$ are components of spin-1/2 operator at site $i$, $\mu$ is the dipole moment on site $i$, and $E$ is the applied electric field. The long-range interaction parameter $J_{ij}$ and the transverse field $\Omega _i$ are selected as 

\begin{equation}
J_{ij}  = \left\{ {\begin{array}{*{20}c}
   {\frac{{J_a }}{{r_{ij}^\sigma  }}\begin{array}{*{20}c}
   {} & {} & {i,j \in A\begin{array}{*{20}c}
   {} & ,  \\
\end{array}}  \\
\end{array}}  \\
   {\frac{{J_b }}{{r_{ij}^\sigma  }}\begin{array}{*{20}c}
   {} & {} & {i,j \in B\begin{array}{*{20}c}
   {} & ,  \\
\end{array}}  \\
\end{array}}  \\
   {\frac{{J_{ab} }}{{r_{ij}^\sigma  }}\begin{array}{*{20}c}
   {} & {} & {i \in A,j \in B,}  \\
\end{array}}  \\
\end{array}} \right.
\end{equation}

\begin{equation}
\Omega _i  = \left\{ {\begin{array}{*{20}c}
   {\Omega _a \begin{array}{*{20}c}
   {} & , & {} & {i \in A\begin{array}{*{20}c}
   {} & ,  \\
\end{array}}  \\
\end{array}}  \\
   {\Omega _b \begin{array}{*{20}c}
   {} & , & {} & {i \in B\begin{array}{*{20}c}
   {} & ,  \\
\end{array}}  \\
\end{array}}  \\
\end{array}} \right.
\end{equation}
where  $J_{a}$, $J_{b}$ and $J_{ab}$ are the nearest neighbor coupling constants in slab A, slab B and between A and B, respectively. $r_{ij}$ is the distance between site $i$ and site $j$. $\sigma $ is introduced to describe the magnitude of the long-range interaction. The case of $\sigma  \to \infty $ corresponds to the weak long-range interaction; the case of $\sigma  \to 0$ corresponds to the strong long-range interaction. In this paper, the interactions of pseudo spins in slab A and slab B as well as the interfacial coupling are all long-range interaction.\\

We assume that the environment of the sites at the same layer is identical, so that the average value of the pseudo spins in the same layer has the same value. We use an improved mean field theory, which take the correlation of the pseudo spins within the range of the eighth neighbor pseudo spins into consideration. That is, within the range of the eighth neighbor pseudo spins, the interaction between pseudo spins is calculated exactly. Within such approximation the average pseudo spin along z direction in the ith layer can be expressed as following: [19,20]

\begin{equation}
R_i  = \left\langle {S_i^z } \right\rangle  = \frac{{\left\langle {H_i^z } \right\rangle }}{{2\left| {H_i } \right|}}\tanh (\frac{{\left| {H_i } \right|}}{{2k_B T}}) , 
\end{equation} 

\begin{equation}
\left\langle {H_i^z } \right\rangle  = \sum\limits_j {J_{ij} } R_j  + 2\mu E , 
\end{equation} 

\begin{equation}
H_i  = \sqrt {\Omega _i^z  + (\left\langle {H_i^z } \right\rangle } )^2 , 
\end{equation} 
where $j$ runs over all layers in one period of the superlattice. In order to make the calculations practicable, the long-range interaction is cut off at the eighth-neighbor in our calculations. The error of this cut-off approximation is negligible, which will be illustrated in sec. III. From Eq. (4), a set of simultaneous nonlinear equations can be obtained. $R_i $ can be calculated numerically.\\

For instance, we take $N_a  = 3$, $N_b  = 3$ :
\[
\left\langle {H_1^z } \right\rangle 
  = \sum\limits_{\scriptstyle i \in 1 \hfill \atop 
  \scriptstyle j \in 1 \hfill} {\frac{{J_a }}{{r_{ij}^\sigma  }}R_1 }  + \sum\limits_{\scriptstyle i \in 1 \hfill \atop 
  \scriptstyle j \in 2 \hfill} {\frac{{J_a }}{{r_{ij}^\sigma  }}R_2 }  + \sum\limits_{\scriptstyle i \in 1 \hfill \atop 
  \scriptstyle j \in 3 \hfill} {\frac{{J_a }}{{r_{ij}^\sigma  }}R_3 }  + \sum\limits_{\scriptstyle i \in 1 \hfill \atop 
  \scriptstyle j \in 4 \hfill} {\frac{{J_{ab} }}{{r_{ij}^\sigma  }}R_4 }  + \sum\limits_{\scriptstyle i \in 1 \hfill \atop 
  \scriptstyle j \in 5 \hfill} {\frac{{J_{ab} }}{{r_{ij}^\sigma  }}R_5 }  + \sum\limits_{\scriptstyle i \in 1 \hfill \atop 
  \scriptstyle j \in 6 \hfill} {\frac{{J_{ab} }}{{r_{ij}^\sigma  }}R_6 }  + 2\mu E
\]
\[
\left\langle {H_2^z } \right\rangle  = \sum\limits_{\scriptstyle i \in 2 \hfill \atop 
  \scriptstyle j \in 1 \hfill} {\frac{{J_a }}{{r_{ij}^\sigma  }}R_1 }  + \sum\limits_{\scriptstyle i \in 2 \hfill \atop 
  \scriptstyle j \in 2 \hfill} {\frac{{J_a }}{{r_{ij}^\sigma  }}R_2 }  + \sum\limits_{\scriptstyle i \in 2 \hfill \atop 
  \scriptstyle j \in 3 \hfill} {\frac{{J_a }}{{r_{ij}^\sigma  }}R_3 }  + \sum\limits_{\scriptstyle i \in 2 \hfill \atop 
  \scriptstyle j \in 4 \hfill} {\frac{{J_{ab} }}{{r_{ij}^\sigma  }}R_4 }  + \sum\limits_{\scriptstyle i \in 2 \hfill \atop 
  \scriptstyle j \in 5 \hfill} {\frac{{J_{ab} }}{{r_{ij}^\sigma  }}R_5 }  + \sum\limits_{\scriptstyle i \in 2 \hfill \atop 
  \scriptstyle j \in 6 \hfill} {\frac{{J_{ab} }}{{r_{ij}^\sigma  }}R_6 }  + 2\mu E
\]
\[
\left\langle {H_3^z } \right\rangle  = \sum\limits_{\scriptstyle i \in 3 \hfill \atop 
  \scriptstyle j \in 1 \hfill} {\frac{{J_a }}{{r_{ij}^\sigma  }}R_1 }  + \sum\limits_{\scriptstyle i \in 3 \hfill \atop 
  \scriptstyle j \in 2 \hfill} {\frac{{J_a }}{{r_{ij}^\sigma  }}R_2 }  + \sum\limits_{\scriptstyle i \in 3 \hfill \atop 
  \scriptstyle j \in 3 \hfill} {\frac{{J_a }}{{r_{ij}^\sigma  }}R_3 }  + \sum\limits_{\scriptstyle i \in 3 \hfill \atop 
  \scriptstyle j \in 4 \hfill} {\frac{{J_{ab} }}{{r_{ij}^\sigma  }}R_4 }  + \sum\limits_{\scriptstyle i \in 3 \hfill \atop 
  \scriptstyle j \in 5 \hfill} {\frac{{J_{ab} }}{{r_{ij}^\sigma  }}R_5 }  + \sum\limits_{\scriptstyle i \in 3 \hfill \atop 
  \scriptstyle j \in 6 \hfill} {\frac{{J_{ab} }}{{r_{ij}^\sigma  }}R_6 }  + 2\mu E
\]
\[
\left\langle {H_4^z } \right\rangle  = \sum\limits_{\scriptstyle i \in 4 \hfill \atop 
  \scriptstyle j \in 1 \hfill} {\frac{{J_{ab} }}{{r_{ij}^\sigma  }}R_1 }  + \sum\limits_{\scriptstyle i \in 4 \hfill \atop 
  \scriptstyle j \in 2 \hfill} {\frac{{J_{ab} }}{{r_{ij}^\sigma  }}R_2 }  + \sum\limits_{\scriptstyle i \in 4 \hfill \atop 
  \scriptstyle j \in 3 \hfill} {\frac{{J_{ab} }}{{r_{ij}^\sigma  }}R_3 }  + \sum\limits_{\scriptstyle i \in 4 \hfill \atop 
  \scriptstyle j \in 4 \hfill} {\frac{{J_b }}{{r_{ij}^\sigma  }}R_4 }  + \sum\limits_{\scriptstyle i \in 4 \hfill \atop 
  \scriptstyle j \in 5 \hfill} {\frac{{J_b }}{{r_{ij}^\sigma  }}R_5 }  + \sum\limits_{\scriptstyle i \in 4 \hfill \atop 
  \scriptstyle j \in 6 \hfill} {\frac{{J_b }}{{r_{ij}^\sigma  }}R_6 }  + 2\mu E
\]
\[
\left\langle {H_5^z } \right\rangle  = \sum\limits_{\scriptstyle i \in 5 \hfill \atop 
  \scriptstyle j \in 1 \hfill} {\frac{{J_{ab} }}{{r_{ij}^\sigma  }}R_1 }  + \sum\limits_{\scriptstyle i \in 5 \hfill \atop 
  \scriptstyle j \in 2 \hfill} {\frac{{J_{ab} }}{{r_{ij}^\sigma  }}R_2 }  + \sum\limits_{\scriptstyle i \in 5 \hfill \atop 
  \scriptstyle j \in 3 \hfill} {\frac{{J_{ab} }}{{r_{ij}^\sigma  }}R_3 }  + \sum\limits_{\scriptstyle i \in 5 \hfill \atop 
  \scriptstyle j \in 4 \hfill} {\frac{{J_b }}{{r_{ij}^\sigma  }}R_4 }  + \sum\limits_{\scriptstyle i \in 5 \hfill \atop 
  \scriptstyle j \in 5 \hfill} {\frac{{J_b }}{{r_{ij}^\sigma  }}R_5 }  + \sum\limits_{\scriptstyle i \in 5 \hfill \atop 
  \scriptstyle j \in 6 \hfill} {\frac{{J_b }}{{r_{ij}^\sigma  }}R_6 }  + 2\mu E
\]
\[
\left\langle {H_6^z } \right\rangle  = \sum\limits_{\scriptstyle i \in 6 \hfill \atop 
  \scriptstyle j \in 1 \hfill} {\frac{{J_{ab} }}{{r_{ij}^\sigma  }}R_1 }  + \sum\limits_{\scriptstyle i \in 6 \hfill \atop 
  \scriptstyle j \in 2 \hfill} {\frac{{J_{ab} }}{{r_{ij}^\sigma  }}R_2 }  + \sum\limits_{\scriptstyle i \in 6 \hfill \atop 
  \scriptstyle j \in 3 \hfill} {\frac{{J_{ab} }}{{r_{ij}^\sigma  }}R_3 }  + \sum\limits_{\scriptstyle i \in 6 \hfill \atop 
  \scriptstyle j \in 4 \hfill} {\frac{{J_b }}{{r_{ij}^\sigma  }}R_4 }  + \sum\limits_{\scriptstyle i \in 6 \hfill \atop 
  \scriptstyle j \in 5 \hfill} {\frac{{J_b }}{{r_{ij}^\sigma  }}R_5 }  + \sum\limits_{\scriptstyle i \in 6 \hfill \atop 
  \scriptstyle j \in 6 \hfill} {\frac{{J_b }}{{r_{ij}^\sigma  }}R_6 }  + 2\mu E
\]

The long-range interaction is cut off at the eighth-neighbor, which indicates that we take $r_{ij} $ as 1, $\sqrt 2 $, $\sqrt 3 $, 2, $\sqrt 5 $, $\sqrt 6 $, $2\sqrt 2 $ and 3. In such approximation, if $\left| {r_i  - r_j } \right| \le 3$, the six above equation will be:

\[
\left\langle {H_1^z } \right\rangle  = \sum\limits_{\scriptstyle i \in 1 \hfill \atop 
  \scriptstyle j \in 1 \hfill} {\frac{{J_a }}{{r_{ij}^\sigma  }}R_1 }  + \sum\limits_{\scriptstyle i \in 1 \hfill \atop 
  \scriptstyle j \in 2 \hfill} {\frac{{J_a }}{{r_{ij}^\sigma  }}R_2 }  + \sum\limits_{\scriptstyle i \in 1 \hfill \atop 
  \scriptstyle j \in 3 \hfill} {\frac{{J_a }}{{r_{ij}^\sigma  }}R_3 }  + 2\sum\limits_{\scriptstyle i \in 1 \hfill \atop 
  \scriptstyle j \in 4 \hfill} {\frac{{J_{ab} }}{{r_{ij}^\sigma  }}R_4 }  + \sum\limits_{\scriptstyle i \in 1 \hfill \atop 
  \scriptstyle j \in 5 \hfill} {\frac{{J_{ab} }}{{r_{ij}^\sigma  }}R_5 }  + \sum\limits_{\scriptstyle i \in 1 \hfill \atop 
  \scriptstyle j \in 6 \hfill} {\frac{{J_{ab} }}{{r_{ij}^\sigma  }}R_6 }  + 2\mu E
\]
\[
\left\langle {H_2^z } \right\rangle  = \sum\limits_{\scriptstyle i \in 2 \hfill \atop 
  \scriptstyle j \in 1 \hfill} {\frac{{J_a }}{{r_{ij}^\sigma  }}R_1 }  + \sum\limits_{\scriptstyle i \in 2 \hfill \atop 
  \scriptstyle j \in 2 \hfill} {\frac{{J_a }}{{r_{ij}^\sigma  }}R_2 }  + \sum\limits_{\scriptstyle i \in 2 \hfill \atop 
  \scriptstyle j \in 3 \hfill} {\frac{{J_a }}{{r_{ij}^\sigma  }}R_3 }  + \sum\limits_{\scriptstyle i \in 2 \hfill \atop 
  \scriptstyle j \in 4 \hfill} {\frac{{J_{ab} }}{{r_{ij}^\sigma  }}R_4 }  + 2\sum\limits_{\scriptstyle i \in 2 \hfill \atop 
  \scriptstyle j \in 5 \hfill} {\frac{{J_{ab} }}{{r_{ij}^\sigma  }}R_5 }  + \sum\limits_{\scriptstyle i \in 2 \hfill \atop 
  \scriptstyle j \in 6 \hfill} {\frac{{J_{ab} }}{{r_{ij}^\sigma  }}R_6 }  + 2\mu E
\]
\[
\left\langle {H_3^z } \right\rangle  = \sum\limits_{\scriptstyle i \in 3 \hfill \atop 
  \scriptstyle j \in 1 \hfill} {\frac{{J_a }}{{r_{ij}^\sigma  }}R_1 }  + \sum\limits_{\scriptstyle i \in 3 \hfill \atop 
  \scriptstyle j \in 2 \hfill} {\frac{{J_a }}{{r_{ij}^\sigma  }}R_2 }  + \sum\limits_{\scriptstyle i \in 3 \hfill \atop 
  \scriptstyle j \in 3 \hfill} {\frac{{J_a }}{{r_{ij}^\sigma  }}R_3 }  + \sum\limits_{\scriptstyle i \in 3 \hfill \atop 
  \scriptstyle j \in 4 \hfill} {\frac{{J_{ab} }}{{r_{ij}^\sigma  }}R_4 }  + \sum\limits_{\scriptstyle i \in 3 \hfill \atop 
  \scriptstyle j \in 5 \hfill} {\frac{{J_{ab} }}{{r_{ij}^\sigma  }}R_5 }  + 2\sum\limits_{\scriptstyle i \in 3 \hfill \atop 
  \scriptstyle j \in 6 \hfill} {\frac{{J_{ab} }}{{r_{ij}^\sigma  }}R_6 }  + 2\mu E
\]
\[
\left\langle {H_4^z } \right\rangle  = 2\sum\limits_{\scriptstyle i \in 4 \hfill \atop 
  \scriptstyle j \in 1 \hfill} {\frac{{J_{ab} }}{{r_{ij}^\sigma  }}R_1 }  + \sum\limits_{\scriptstyle i \in 4 \hfill \atop 
  \scriptstyle j \in 2 \hfill} {\frac{{J_{ab} }}{{r_{ij}^\sigma  }}R_2 }  + \sum\limits_{\scriptstyle i \in 4 \hfill \atop 
  \scriptstyle j \in 3 \hfill} {\frac{{J_{ab} }}{{r_{ij}^\sigma  }}R_3 }  + \sum\limits_{\scriptstyle i \in 4 \hfill \atop 
  \scriptstyle j \in 4 \hfill} {\frac{{J_b }}{{r_{ij}^\sigma  }}R_4 }  + \sum\limits_{\scriptstyle i \in 4 \hfill \atop 
  \scriptstyle j \in 5 \hfill} {\frac{{J_b }}{{r_{ij}^\sigma  }}R_5 }  + \sum\limits_{\scriptstyle i \in 4 \hfill \atop 
  \scriptstyle j \in 6 \hfill} {\frac{{J_b }}{{r_{ij}^\sigma  }}R_6 }  + 2\mu E
\]
\[
\left\langle {H_5^z } \right\rangle  = \sum\limits_{\scriptstyle i \in 5 \hfill \atop 
  \scriptstyle j \in 1 \hfill} {\frac{{J_{ab} }}{{r_{ij}^\sigma  }}R_1 }  + 2\sum\limits_{\scriptstyle i \in 5 \hfill \atop 
  \scriptstyle j \in 2 \hfill} {\frac{{J_{ab} }}{{r_{ij}^\sigma  }}R_2 }  + \sum\limits_{\scriptstyle i \in 5 \hfill \atop 
  \scriptstyle j \in 3 \hfill} {\frac{{J_{ab} }}{{r_{ij}^\sigma  }}R_3 }  + \sum\limits_{\scriptstyle i \in 5 \hfill \atop 
  \scriptstyle j \in 4 \hfill} {\frac{{J_b }}{{r_{ij}^\sigma  }}R_4 }  + \sum\limits_{\scriptstyle i \in 5 \hfill \atop 
  \scriptstyle j \in 5 \hfill} {\frac{{J_b }}{{r_{ij}^\sigma  }}R_5 }  + \sum\limits_{\scriptstyle i \in 5 \hfill \atop 
  \scriptstyle j \in 6 \hfill} {\frac{{J_b }}{{r_{ij}^\sigma  }}R_6 }  + 2\mu E
\]
\[
\left\langle {H_6^z } \right\rangle  = \sum\limits_{\scriptstyle i \in 6 \hfill \atop 
  \scriptstyle j \in 1 \hfill} {\frac{{J_{ab} }}{{r_{ij}^\sigma  }}R_1 }  + \sum\limits_{\scriptstyle i \in 6 \hfill \atop 
  \scriptstyle j \in 2 \hfill} {\frac{{J_{ab} }}{{r_{ij}^\sigma  }}R_2 }  + 2\sum\limits_{\scriptstyle i \in 6 \hfill \atop 
  \scriptstyle j \in 3 \hfill} {\frac{{J_{ab} }}{{r_{ij}^\sigma  }}R_3 }  + \sum\limits_{\scriptstyle i \in 6 \hfill \atop 
  \scriptstyle j \in 4 \hfill} {\frac{{J_b }}{{r_{ij}^\sigma  }}R_4 }  + \sum\limits_{\scriptstyle i \in 6 \hfill \atop 
  \scriptstyle j \in 5 \hfill} {\frac{{J_b }}{{r_{ij}^\sigma  }}R_5 }  + \sum\limits_{\scriptstyle i \in 6 \hfill \atop 
  \scriptstyle j \in 6 \hfill} {\frac{{J_b }}{{r_{ij}^\sigma  }}R_6 }  + 2\mu E
\]

From the above equations, we can see that the mean field (or local field) of pseudo spins is related to the mean value of $R_1 $, $R_2 $, $R_3 $, $R_4 $, $R_5 $ and $R_6 $ (according to our cut-off approximation). Therefore, the correlation between the pseudo spins has been partially considered in the range of the eighth-neighbor pseudo spins in our calculation.\\

The polarization of the $ith$ layer in one period of the superlattice is 
\begin{equation}
P_i  = 2n\mu R_i ,
\end{equation}
where $n$ is the number of pseudo spins in a unit volume. 
The mean polarization of the superlattice can be given as:
\begin{equation}
\overline P  = \frac{1}{N}\sum\limits_{i = 1}^N {P_i } , 
\end{equation}
where $N$ is total layer-numbers of the two materials in one period of the superlattice.\\

The pyroelectric coefficient of the $ith$ layer can be expressed by:

\begin{equation}
p_i  = - \frac{{\partial P_i }}{{\partial T}} = - 2n\mu \frac{{\partial R_i }}{{\partial T}}, 
\end{equation}
Then, the average pyroelectric coefficient of the ferroelectric superlattice is:

\begin{equation}
p = - \frac{{\partial \overline P }}{{\partial T}} = - \frac{1}{N}\sum\limits_{i = 1}^N {\frac{{\partial P_i }}{{\partial T}}}  = - \frac{1}{N}\sum\limits_{i = 1}^N {2n\mu \frac{{\partial R_i }}{{\partial T}}}, 
\end{equation}

The deviations $\frac{{\partial R_i }}{{\partial T}}$ can be obtained by numerically differential calculation. Then the pyroelectric coefficient is obtained numerically. By changing the values of $\sigma $ and $J_{ab} $, the effects of the long-range interaction and the interfacial coupling on the pyroelectric coefficient of the ferroelectric superlattice are investigated.\\

{\bf{III. Results AND Discussions}}\\

Firstly, in order to make the calculations practicable, we will restrict the infinite long-range interaction to a finite range. In Fig. 2, we give the polarization of the ferroelectric superlattice as a function of the temperature at different cut-off approximations. The parameters selected in Fig. 2 are $N_a  = N_b  = 3$, $\Omega _a /J_b  = \Omega _b /J_b  = 1.0$, $J_{ab}  = (J_a  + J_b)/2$, $\sigma  = 3.0$, $J_b $ is taken as the unit of the energy and $J_a $ is fixed as $2J_b$ in the whole paper. From Fig. 2, the more neighbors we consider, the smaller error is. We can see that the error is negligible if the long-range interaction is cut off at the eighth-neighbor. So, the long-range interaction will be cut off at the eighth-neighbor in the following calculations. \\  

The effects of the long-range interaction on the pyroelectric coefficient of the superlattice are shown in Fig. 3, the parameters are selected as $N_a  = N_b  = 5$, $\Omega _a /J_b  = \Omega _b /J_b  = 1.0$, $J_{ab}  = (J_a  + J_b)/2$. It is found that the pyroelectric coefficient decreases with the increase of the long-range interaction when the temperature is lower than the phase transition temperature. From Fig. 3, we can obtain that when the $\sigma  \to \infty $, two peaks of the pyroelectric coefficient occur, one is low and round; the other is high and sharp. The lower pyroelectric peak is mainly caused by occurrence of the paraelectric phase in the material B. Under the influence of the material A, the pyroelectric properties of the material B in the superlattice is different from the single bulk B, of which the pyroelectric peak is a sharp one at the phase transition temperature of the bulk B. As the strength of the long-range interaction increasing, the lower pyroelectric peak will become more and more round until disappear. Simultaneously, we also obtain that the pyroelectric peaks shift towards the region of high temperature. It means that the phase transition temperature increases with the increase of the long-range interaction.  The system behaves as a single material because of the strong long-range interaction.\\

The effects of the thickness of one period on the pyroelectric coefficients of the superlattice are given in Fig. 4. In Fig. 4 (a), we consider the case of the strong long-range interaction, the parameters are selected as $\Omega _a /J_b  = \Omega _b /J_b  = 1.0$, $J_{ab}  = (J_a  + J_b)/2$, $\sigma  = 2.0$. In Fig. 4 (b), we consider the case of the weak long-range interaction, the parameters are selected as $\Omega _a /J_b  = \Omega _b /J_b  = 1.0$, $J_{ab}  = (J_a  + J_b)/2$, $\sigma  \to \infty $. From the two figures, we can find some similar phenomena. Firstly, the phase transition temperature increase when the thickness of one period become thicker. Secondly, there are the two pyroelectric peaks when the thickness of one period is thick enough. The lower pyroelectric peak is lead to by the appearance of the paraelectric phase in the material B. Comparing the two figures; we can see some obviously different phenomena. Firstly, for the superlattice ($N_a  = N_b  = 3$), which the thickness of one period is thin, the long-range interaction will cause some pyroelectric peaks disappear. In Fig. 4 (a), there is only one sharp peak for the strong long-range interaction ($\sigma  = 2.0$). The results show that when the thickness of one period of the ferroelectric superlattice is thin, the long-range interaction has an important role on the number of the pyroelectric peaks. We think that the strong long-range interaction makes material A, material B combined into a whole body, which will contribute to a single effect. From Fig. 4 (b), we can obtain that when only consider the weak long-range interaction, there are still two pyroelectric peaks, one is low and round and the other is high and sharp. When the thickness of one period is thick enough, the long-range interaction has trivial influence on the number of the pyroelectric peaks; there are still two sharp peaks. Secondly, the change of the phase transition temperature dependent of the thickness of one period is obvious when we consider the strong long-range interaction.\\ 

In Fig. 5, we give the effects of the interfacial coupling on the pyroelectric coefficient of the superlattice ($N_a  = N_b  = 10$). In Fig. 5 (a), we consider the case of the strong long-range interaction, the parameters are selected as $\Omega _a /J_b  = \Omega _b /J_b  = 2.0$, $\sigma  = 2.0$. In Fig. 5 (b), we consider the case of the weak long-range interaction, the parameters are selected as $\Omega _a /J_b  = \Omega _b /J_b  = 2.0$, $\sigma  \to \infty $. From the two figures, we can obtain the three pyroelectric peaks when the interfacial coupling is strong. The phenomenon that the number of the pyroelectric peaks increase is mainly caused by the strong interfacial coupling. With the decrease of the interfacial coupling, the peak of the interfacial pyroelectric coefficient is merged into the peaks of the pyroelectric coefficient of slab A and slab B. So the three peaks will become two peaks. We can also obtain that with the increase of the interfacial coupling, the phase transition temperature of the superlattice will increase and the pyroelectric coefficient decreases before the phase transition. But there are some obvious differences in the two graphs. The three pyroelectric peaks in Fig. 5 (a) are all sharper than those in Fig. 5 (b) under the same condition ($J_{ab} /J_b  = 8.0$). Thus, we can conclude that with the increase of the long-range interaction, some pyroelectric peaks disappear gradually. Comparing the two graphs, we also obtain that in Fig. 5 (a), the strong long-range interaction is considered, the phenomenon which the pyroelectric coefficient decrease before the phase transition with the increase of the interfacial coupling is obvious; in Fig. 5 (b), only the weak long-range interaction is considered, the phenomenon is not so obvious.\\  
 
In summary, the effects of the long-range interaction on the pyroelectric properties of the ferroelectric superlattice are investigated. In the meantime, we also study the effects of the interfacial coupling and the thickness of one period on the pyroelectric coefficient by taking into account the long-range interaction.  From the above results, we obtain that: (1) With the increase of the long-range interaction, the pyroelectric coefficient decreases before the phase transition and the phase transition temperature of the ferroelectric superlattice increases. (2) In the framework of considering the long-range interaction, the pyroelectric coefficient decreases before the phase transition with the increase of the interfacial coupling, and the phase transition temperature increases with the increase of the interfacial coupling and the thickness of one period. (3) For the strong interfacial coupling, there exist three pyroelectric peaks, with the decrease of the interfacial coupling, the three peaks will become two gradually. (4) With the decrease of the thickness of one period, the two pyroelectric peaks become one gradually.\\

{\bf {Acknowledgments:}}

Project was supported by the National Natural Science Foundation of China (Grant No.10174049) and the Natural Science Foundation of JiangSu Education Committee of China (Grant No.00KJB140009).

\newpage
\begin{center}{CAPTION OF FIGURES}\end{center}

Fig. 1:\\
The schematic of a ferroelectric superlattice composed of two different ferroelectric slabs in one period.\\ 

Fig. 2:\\
The polarization of the superlattice as a function of temperature for different cut-off approximations.\\   

Fig. 3:\\
The pyroelectric coefficient of the superlattice ($N_a  = N_b  = 5$) as a function of temperature for different strength of long-range interaction. $\Omega _a /J_b  = \Omega _b /J_b  = 1.0$, $J_{ab}  = (J_a  + J_b)/2$, $J_a  = 2J_b $.\\

Fig. 4:\\
The pyroelectric coefficient of the superlattice as a function of temperature for different thickness of one period by taking into account the long-range interaction. $\Omega _a /J_b  = \Omega _b /J_b  = 1.0$. $J_{ab}  = (J_a  + J_b)/2$, $J_a  = 2J_b $. (a): $\sigma = 2.0 $; (b): $\sigma  \to \infty $.\\

Fig. 5:\\
The pyroelectric coefficient of the superlattice ($N_a  = N_b  = 10$) as a function of temperature for different strength of interfacial coupling by taking into account the long-range interaction. $\Omega _a /J_b  = \Omega _b /J_b  = 2.0$, $J_a  = 2J_b $. (a): $\sigma = 2.0 $; (b): $\sigma  \to \infty $.

\newpage
\vfil\includegraphics[scale=0.7]{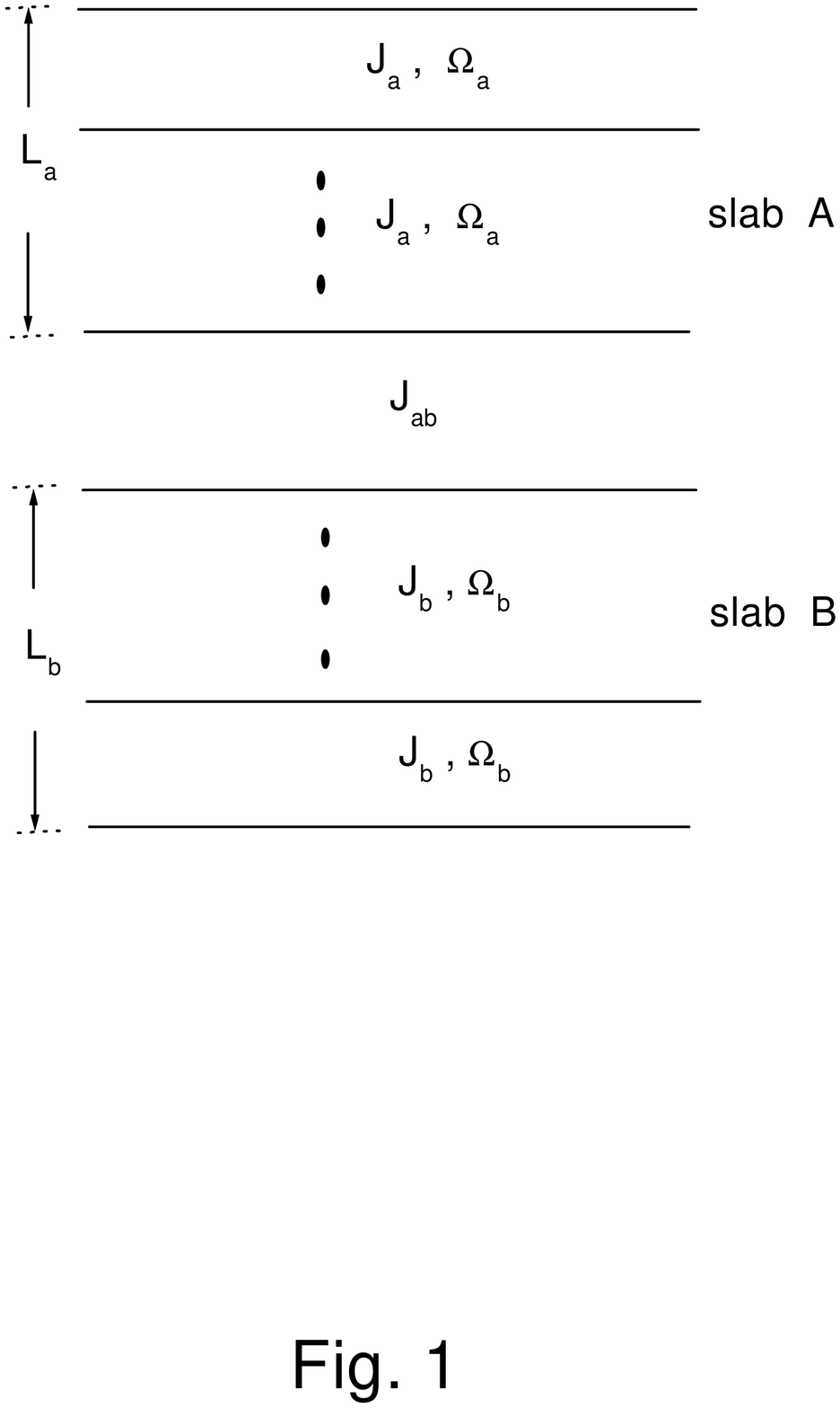}\vfil

\newpage
\vfil\includegraphics[scale=0.7]{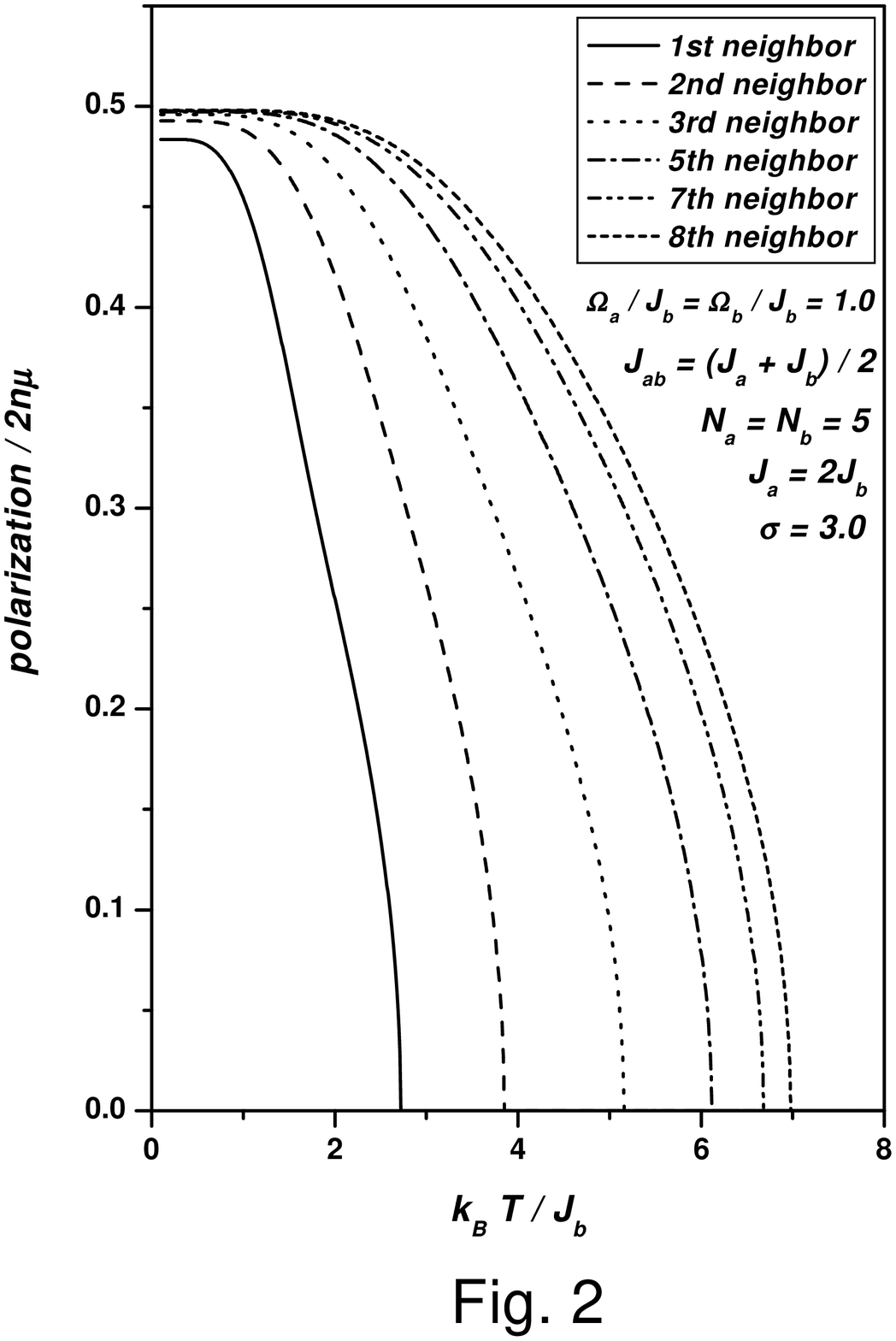}\vfil

\newpage
\vfil\includegraphics[scale=0.7]{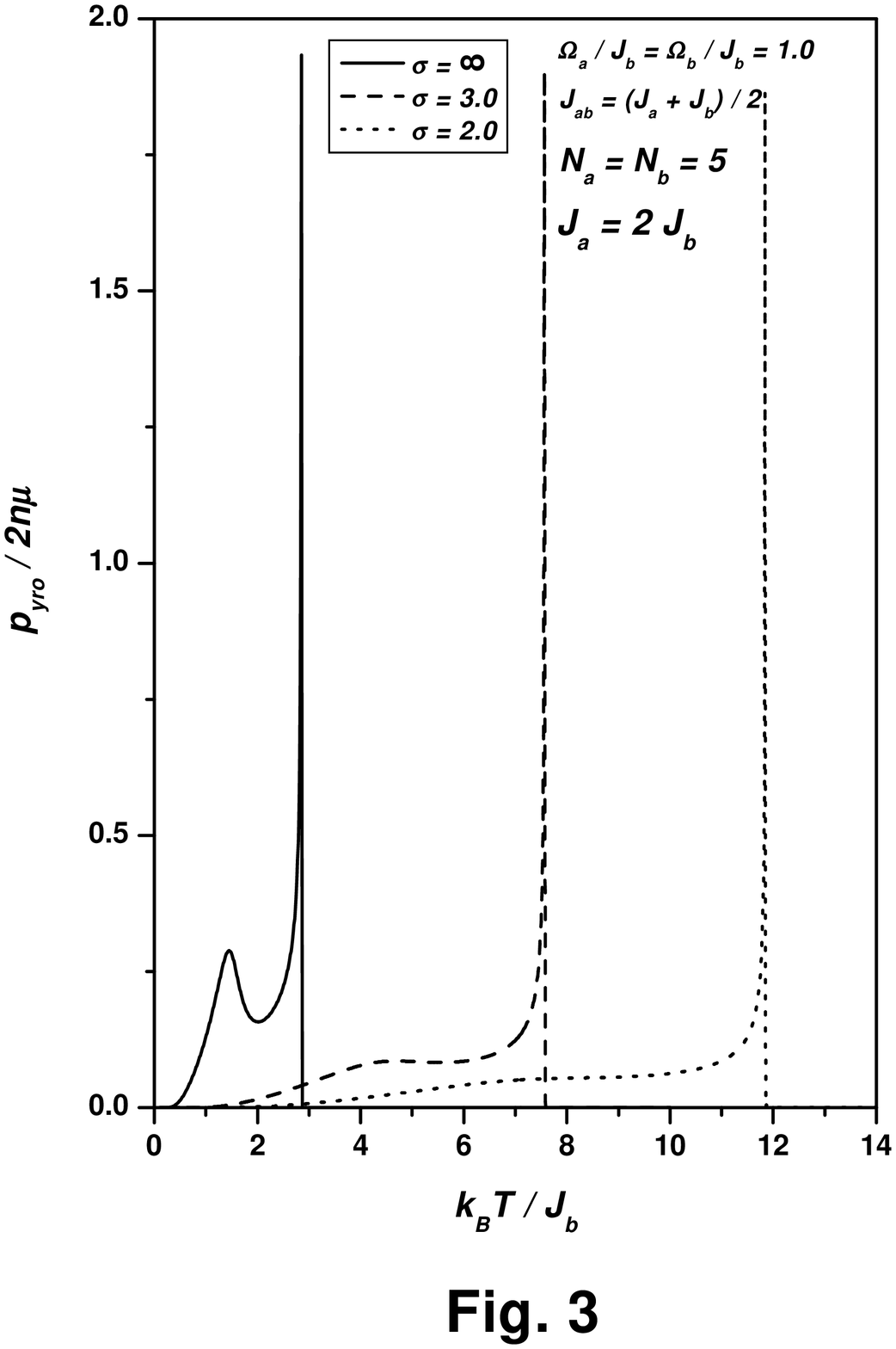}\vfil

\newpage
\vfil\includegraphics[scale=0.7]{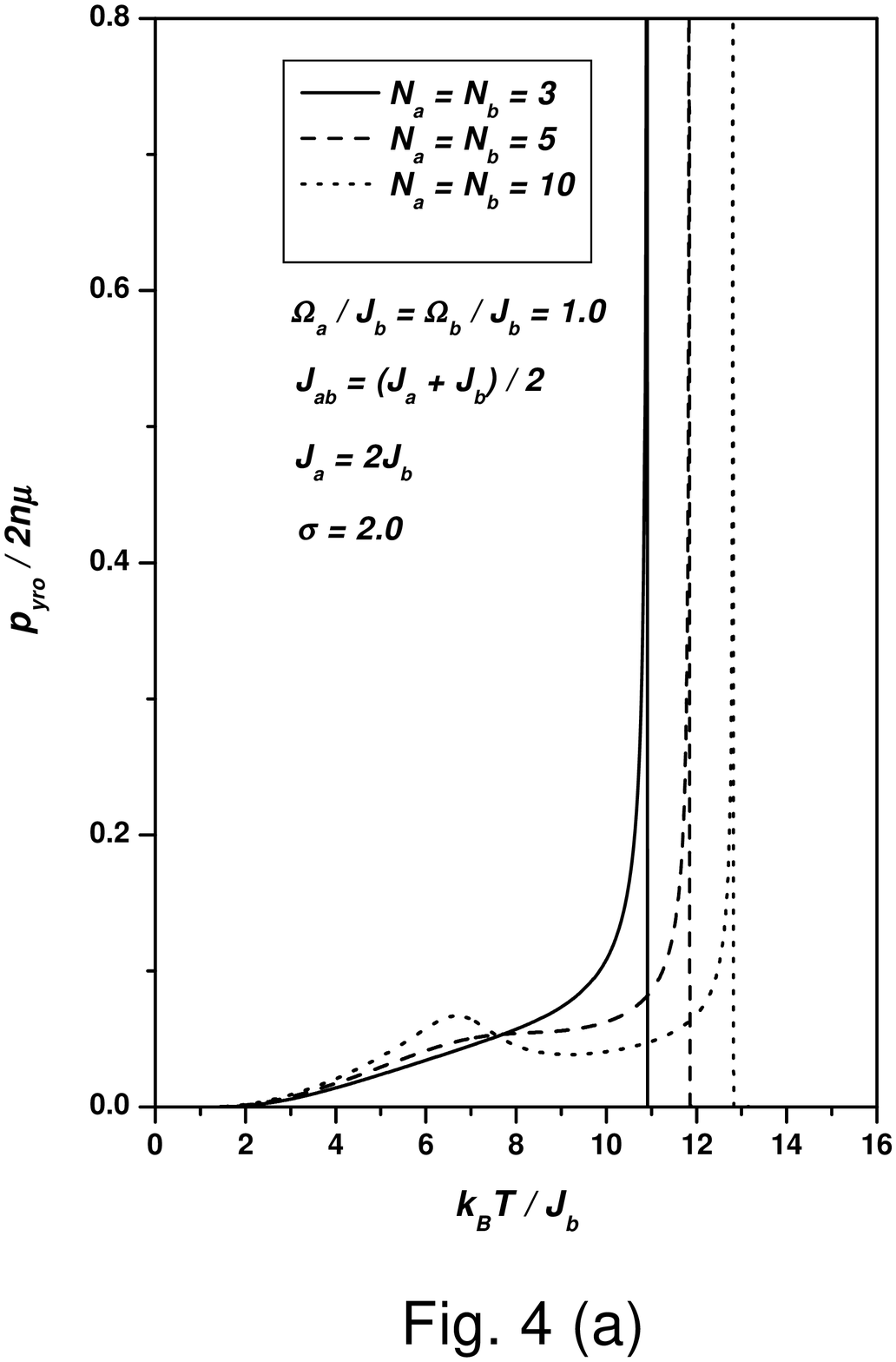}\vfil

\newpage
\vfil\includegraphics[scale=0.7]{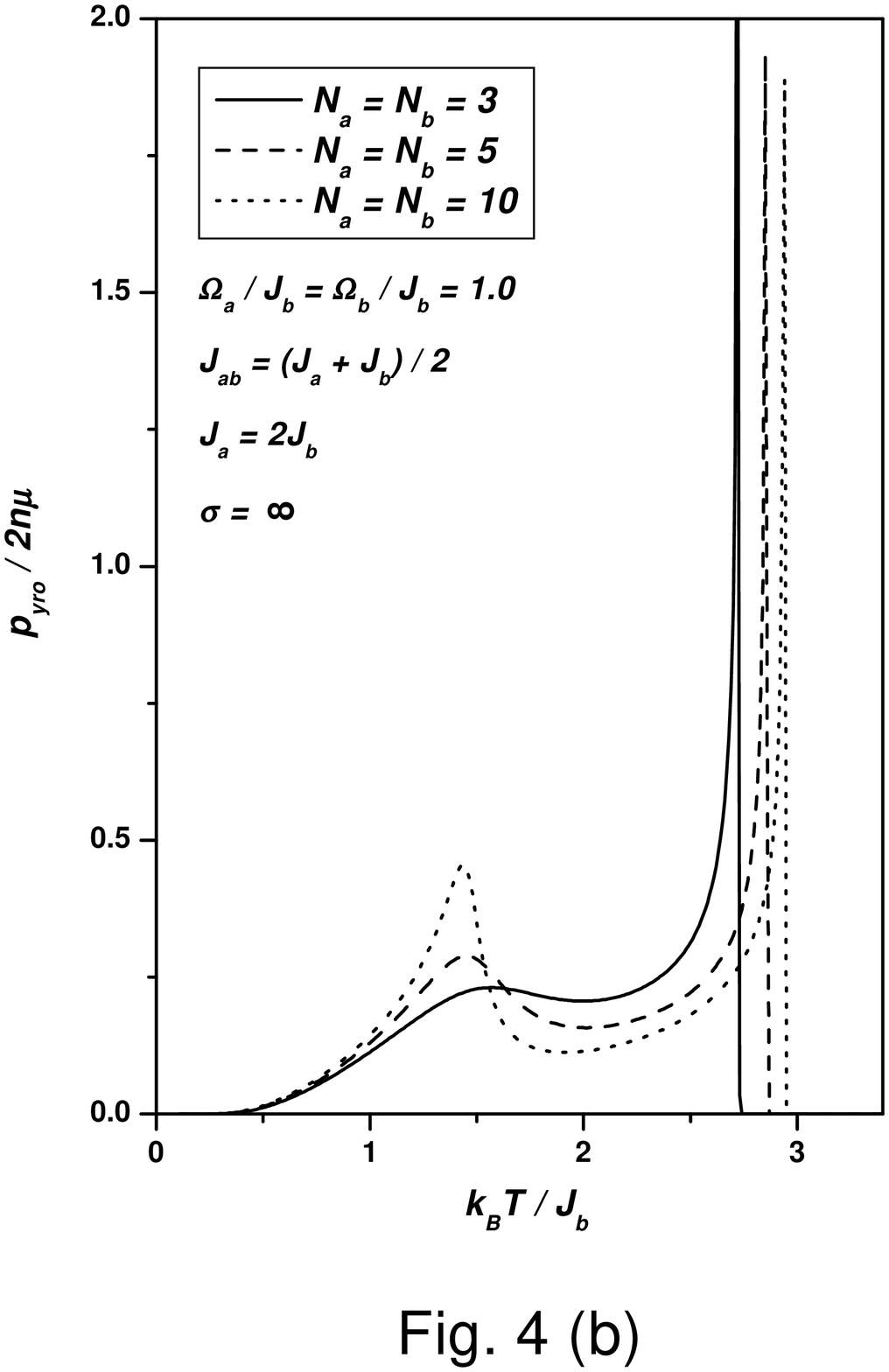}\vfil

\newpage
\vfil\includegraphics[scale=0.7]{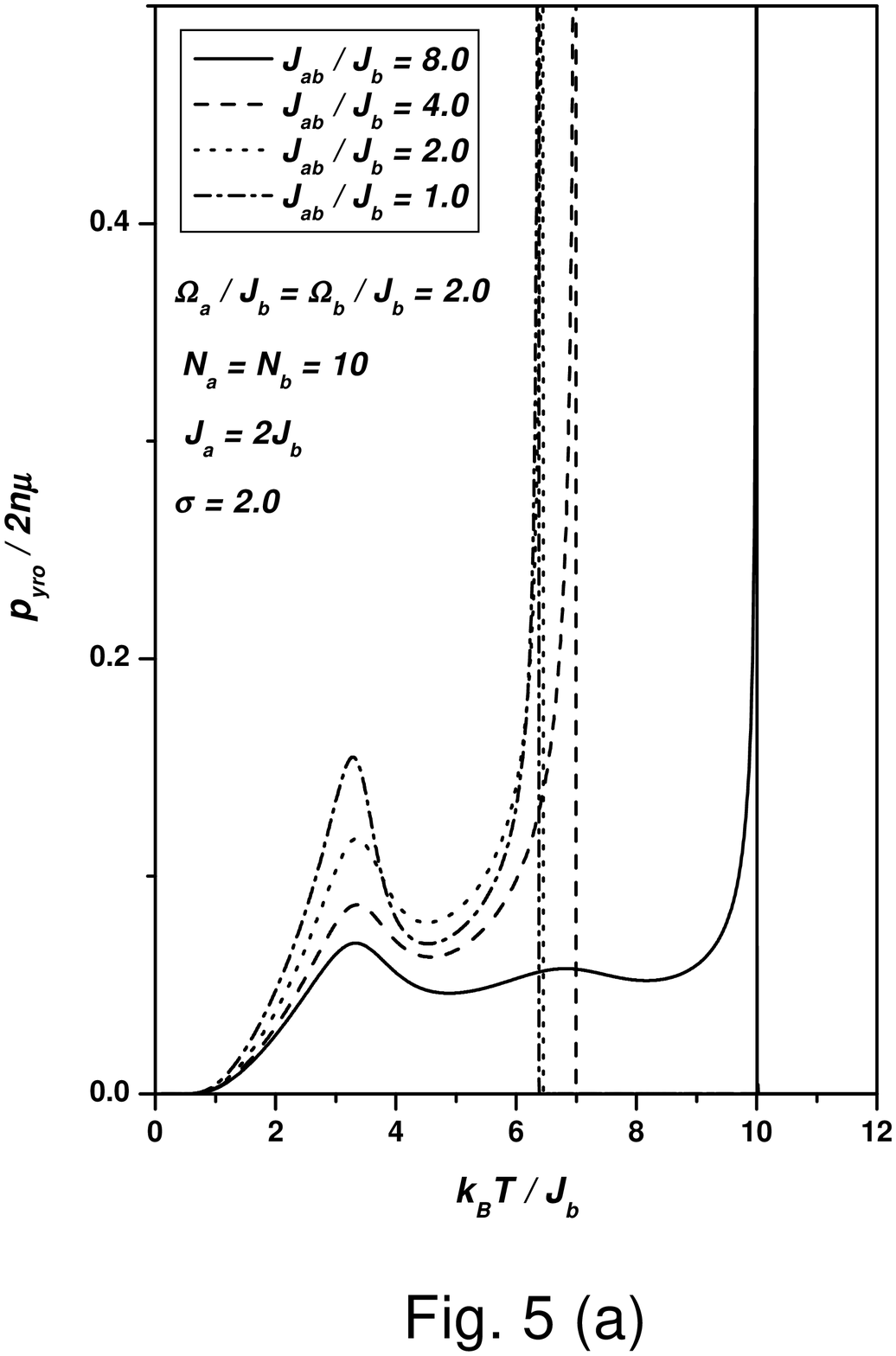}\vfil

\newpage
\vfil\includegraphics[scale=0.7]{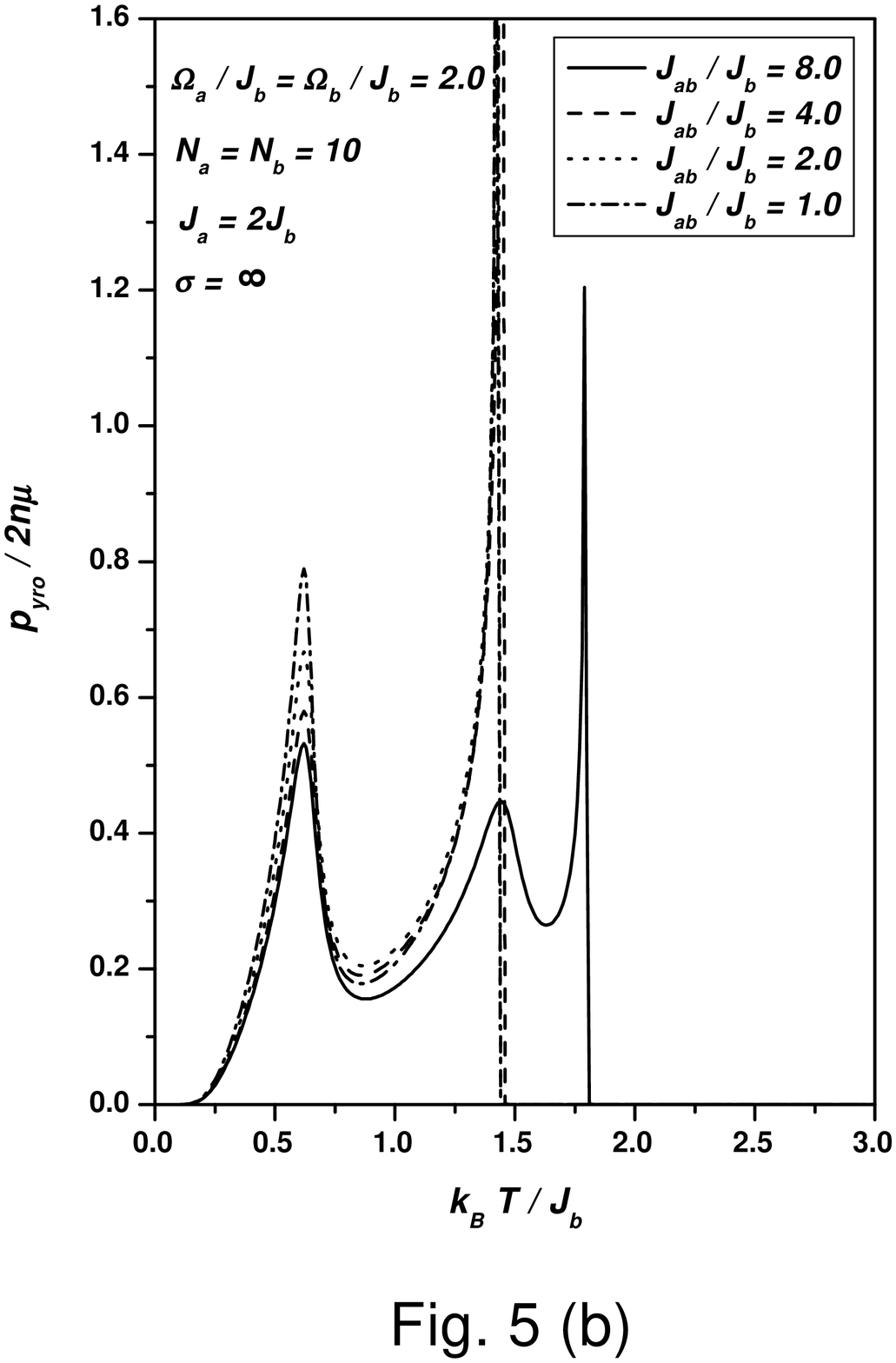}\vfil

\end{document}